\documentclass[12pt]{iopart}

\usepackage{epsfig}
\def\lsim{\mathrel{\raise.3ex\hbox{$<$\kern-.75em\lower1ex\hbox{$\sim$}}}}
\def\gsim{\mathrel{\raise.3ex\hbox{$>$\kern-.75em\lower1ex\hbox{$\sim$}}}}

\def\Mp{{M_{\rm P}}}

\begin{document}
\hspace{12.3cm}
\normalsize DESY 07-128\hfill\mbox{}

\title{Gamma-Rays from Decaying Dark Matter}
\author{Gianfranco Bertone$^1$, Wilfried Buchm\"uller $^2$, 
Laura Covi $^2$, Alejandro Ibarra $^2$}
\address{$^1$ Institut d'Astrophysique de Paris, 
UMR 7095-CNRS,Universit\'e Pierre et Marie Curie, 
98bis boulevard Arago, 75014 Paris, France \\
$^2$ Deutsches Elektronen-Synchrotron DESY, Hamburg, Germany} 
%\ead{graham.douglas@iop.org}
\begin{abstract}
We study the prospects for detecting gamma-rays from decaying Dark Matter
(DM), focusing in particular on gravitino DM in R-parity breaking vacua. 
Given the substantially different angular distribution of the predicted 
gamma-ray signal with respect to the case of annihilating DM, and the 
relatively poor (of order 0.1$^\circ$) angular resolution of gamma-ray 
detectors, the best strategy for detection is in this case to look for 
an exotic contribution to the gamma-ray flux at high galactic latitudes, 
where the decaying DM contribution would resemble an astrophysical 
extra-galactic component, similar to the one inferred by EGRET 
observations. Upcoming experiments such as GLAST and AMS-02
may identify this exotic contribution and discriminate
it from astrophysical sources, or place significant constraints
on the mass and lifetime of DM particles. 
\end{abstract}

\section{Introduction}
A tremendous theoretical and experimental effort is in progress to 
clarify the nature of the elusive {\it Dark Matter} that appears to 
dominate the matter density of the 
Universe~\cite{Bertone:2004pz,Bergstrom:2000pn}.
The most studied DM candidates are Weakly Interacting Massive Particles, 
that achieve the appropriate relic density by {\it freezing-out} of 
thermal equilibrium when their self-annihilation rate becomes smaller 
than the expansion rate of the Universe. 
The characteristic mass of these particles is $\cal{O}$$(100)$ GeV, and
the most representative and commonly discussed candidates in this class 
of models are the supersymmetric neutralino, and the B$^{(1)}$ particle, 
first excitation of the hypercharge gauge boson, in theories with 
Universal Extra Dimensions.
These particle are and will be searched for via collider, direct and 
indirect searches. In particular, the latter are based on the very same 
mechanism that controls the relic density of DM, i.e., 
self-annihilations. 
In fact, although the annihilation rate in the local Universe is on 
average severely suppressed, it can still be extremely high at the 
centre of dense DM halos, since it is proportional to the square of the 
DM particles number density. The prospects for indirect detection of 
annihilating DM have been extensively discussed 
(see ~\cite{Bertone:2004pz,Bergstrom:2000pn} and references therein).

However, self-annihilating relics are not the only DM candidates, and 
indirect DM searches are not only relevant for self-annihilating 
particles. Three of us have recently studied an excellent DM candidate, 
the gravitino in R-parity breaking vacua, that can achieve the 
appropriate relic density through the thermal production in the early 
high-temperature phase of the Universe, and that naturally leads to 
a cosmological history consistent with thermal leptogenesis and 
primordial nucleosynthesis~\cite{bcx07}.

Since R-parity is broken, gravitinos can decay into a photon and 
a neutrino~\cite{ber91}, although with a lifetime that, being 
suppressed both by the Planck mass and by the small R-parity breaking 
parameters, is naturally much longer than the age of the 
Universe~\cite{ty00}. Similarly, for sufficiently small R-parity 
breaking, also neutralinos~\cite{bs87,bb88} and axinos~\cite{kk02} 
are dark matter candidates which can decay into a photon and a neutrino. 
These scenarios thus predict a diffuse flux of photons and 
neutrinos that, by comparison with existing observational data, can 
be used to set constraints on the mass and lifetime of the decaying 
particles. 
Interestingly, an excess in the galactic component of the diffuse 
gamma-ray flux measured by EGRET has been claimed in 
Ref.~\cite{Sreekumar:1997un}, at energies between 1 and 10 GeV.
A more careful analysis of the Galactic foreground has led 
Strong~{\it et al.} 
to a new estimate of the extra-galactic component~\cite{Strong:2004ry}
with a significantly different spectrum with respect to the previous
analysis. More recently, Stecker {\it et al.}~\cite{Stecker:2007xp} 
pointed out a possible error in the energy calibration of EGRET 
above 1 GeV, a circumstance that if confirmed would make any 
interpretation of EGRET data in terms of exotic components, 
such as DM annihilation or decay, unreliable, if not meaningless.

In view of these and other systematic uncertainties \cite{mxx06},
we will not try here to fit 
the EGRET data with the gamma-ray flux produced by decaying DM, although 
we regard this coincidence as interesting and deserving further 
attention. We perform instead a careful analysis of the signal that 
might be detected with the next generation of gamma-ray experiments.
%, in particular with the upcoming space satellite GLAST. 
Similar analyses has previously been carried out for decaying DM 
candidates with masses in the keV range, such as a scalar 
modulus~\cite{ahx98} or a sterile neutrino~\cite{De Rujula:1980qd,bnx06}.
More recently also the case of small mass splittings and heavy
DM decaying into MeV photons has been discussed in order to explain 
the COMPTEL excess in the photon flux~\cite{Cembranos:2007fj,Olive:1985bd}.

Unlike the case of stable neutralinos and other WIMPs, the rate at which 
gravitinos produce photons is proportional to the density of DM 
particles, as appropriate for decaying DM particles, not to the 
{\it square} of the DM density. As a consequence, the strategies 
for indirect detection must keep into account the different angular 
distribution of the predicted signal, and the different ratio between 
galactic and extra-galactic contributions. 
Although the situation is very similar to the case of other decaying DM 
candidates, such as decaying sterile neutrinos, the angular resolution 
of experiments sensitive to photons from decaying gravitinos, typically 
above 5 GeV~\cite{bcx07}, are much worse than X-ray telescopes, 
relevant for sterile neutrinos. Here we study the best strategies to 
detect an exotic component in the gamma-ray diffuse flux with future 
experiments such as the upcoming gamma-ray satellite GLAST, scheduled 
for launch in the next few months, and with AMS-02.
Although in our analysis we adopt gravitinos as our fiducial DM 
candidates, our results can be applied to any decaying DM particle 
in a similar range of masses into monochromatic photons.

The paper is organised as follows: in Sec.~2 we discuss our expectation 
for gravitino lifetime and the decay channel, while in Sec.~3 the 
Galactic and extra-galactic contributions to the gamma-ray 
flux from gravitino decays. In Sec.~4 we consider alternative
targets for indirect detection such as dwarf galaxies and galaxy 
clusters. In Sec.~5 we discuss our results and compare the indirect 
detection strategies of decaying and self-annihilating
DM. Finally we give our conclusions in Sec.~6.

\section{Gravitino decay}

As is the case for the proton, we do not know if the DM particle is 
absolutely stable. 
In the case of supersymmetric candidates, usually R-parity is invoked to 
make the proton sufficiently long-lived and it automatically gives 
that the Lightest Supersymmetric Particle (LSP) is stable. 
On the other hand R-parity is not the only possibility for protecting
the proton from rapid decay and a small amount of R-parity violation
does not rule out the possibility of having supersymmetric DM if the
LSP is very weakly interacting with the R-parity violating sector
or the decay is highly suppressed by phase space.
As an example, in \cite{bcx07}, we have considered a supersymmetric 
extension of the standard model with small R-parity and lepton number 
violating couplings and a gravitino LSP. 
The model predicts a small photino-neutrino mixing  
$|U_{\tilde\gamma\nu}|={\cal O}(10^{-8})$, which leads to the decay of
the gravitino into photon and neutrino \cite{ty00},
\begin{eqnarray}
\label{gamma-gravitino}
\Gamma (\psi_{3/2} \rightarrow \gamma \nu)
= {1\over 32 \pi} |U_{\tilde\gamma\nu}|^2 \frac{m_{3/2}^3}{\Mp^2}\; .
\end{eqnarray}
Using $\Mp = 2.4 \times 10^{18} $ GeV, one obtains for
the gravitino lifetime 
\begin{eqnarray}
\label{grav-lifetime}
\tau^{\rm 2-body}_{3/2} \simeq  3.8\times 10^{27} {\rm s} 
\left(\frac{|U_{\tilde\gamma\nu}|}{10^{-8}}\right)^{-2}
\left(\frac{m_{3/2}}{10 \mbox{ GeV}} \right)^{-3}\; .
\end{eqnarray}

At tree level this decay channel can be suppressed if the sneutrino 
$v.e.v.$ responsible for the photino-neutrino mixing is very small, 
but even if the mixing vanishes, the decay can take place via one 
loop diagrams. The loop induced decay has been recently computed 
in~\cite{lor07}, where it has been shown that also in this case the 
channel dominates over the 3-body decay into 
fermions~\cite{Moreau:2001sr}
for small gravitino masses. For this reason we will concentrate on this 
particular channel and will assume in the following that our 
Dark Matter candidate decays into a photon and neutrino producing 
two monochromatic lines at energy equal to 
$m_{DM}/2$ with a lifetime of the order of $10^{27}{\rm s}$ or larger.
On the other hand, if the gravitino is sufficiently heavy
it could decay into $W$ or $Z$ bosons, producing through
fragmentation a continuous spectrum of photons with a
characteristic shape~\cite{Ibarra-Tran}.
The neutrino flux in the few GeVs energy range is unfortunately 
overwhelmed by the atmospheric neutrino background and so its detection 
seems much more difficult than that of the gamma-ray flux.

Note that the signal in gammas would be the same, only twice as strong, 
for the case of a scalar DM candidate decaying into two photons.

\section{Gamma-Rays from DM decay}

If the DM particles decay all around us, we expect two sources for
a diffuse background. We have the DM decaying in the Milky Way halo
nearby and in addition those decaying at cosmological distances.

Let us first consider the latter ones, which have been more intensively 
studied in the literature \cite{ty00,Overduin:2004sz}. 
The decay of DM into photon and neutrino at cosmological distances 
gives rise to a perfectly isotropic extragalactic diffuse gamma-ray 
flux with a characteristic energy spectrum, corresponding to a 
red--shifted monochromatic line. 
A photon with measured energy $E = m_{3/2}/(2(1+z))$ has been 
emitted at the comoving distance $\chi(z)$, with 
$d\chi/dz=(1+z)^{-3/2}/(a_0 H_0\sqrt{\Omega_M(1+ \kappa (1+z)^{-3})})$. 
Here $a_0$ and $H_0$ are the present scale factor and Hubble parameter, 
respectively, and $\kappa = \Omega_\Lambda/\Omega_M \simeq 3$, with 
$\Omega_\Lambda + \Omega_M =1$, assuming a flat universe.
Then we obtain for the photon flux 
\begin{equation}
\label{photon-flux}\label{signal}
{dJ_{eg}\over dE} = A_{eg} \frac{2}{m_{DM}}
\left(1+ \kappa \left({2E\over m_{DM}}\right)^3\right)^{-1/2} 
\left({2E\over m_{DM}}\right)^{1/2} 
\Theta\left(1 - \frac{2E}{m_{DM}}\right)\; , 
\end{equation}
with 
\begin{equation}\label{cgamma}
A_{eg} = 
\frac{\Omega_{DM} \rho_c}{4\pi\tau_{DM} m_{DM} H_0 \Omega_M^{1/2}}
= 10^{-7}\ ({\rm cm}^2{\rm s}~{\rm str})^{-1} 
\left(\frac{\tau_{DM}}{ 10^{27}\ \mbox{s}} \right)^{-1}
\left(\frac{m_{DM}}{ 10\ \mbox{GeV}} \right)^{-1};
\end{equation}
here $ \tau_{DM} $ is the DM particle lifetime and is 
given by Eq.~(\ref{grav-lifetime}) for the gravitino case.
We have taken the particle density to be equal to the Cold Dark 
Matter density as $\Omega_{DM} h^2 =0.1$, and the other constants
as $\rho_c = 1.05\; h^2 \times 10^{-5} {\rm GeV} {\rm cm}^{-3}$,
total matter density $\Omega_M = 0.25$ and
$H_0 = h\; 100\; {\rm km}\; {\rm s}^{-1}\; {\rm Mpc}^{-1}$ 
with $h=0.73$~\cite{Yao:2006px}. 
We are considering here $\tau_{DM} \gg H_0^{-1} $ so that we
can neglect in the above formula the depletion of the number 
density due to the decay.

In addition to the extragalactic signal there is an anisotropic sharp 
line from the halo of our galaxy with an intensity comparable to the 
extragalactic signal~\cite{ahx98}.
The flux from the decay of halo DM particles is given by the
density profile, i.e.
\begin{equation}
\label{photon-flux-halo}\label{haloflux}
{dJ_{halo}\over dE} = A_{halo}\; \frac{2}{m_{DM}} 
%\left({2E\over m_{DM}}\right)^2
\delta\left(1 - \frac{2E}{m_{DM}}\right)\; , 
\end{equation}
where
\begin{equation}\label{dgamma}
A_{halo} = 
\frac{1}{4\pi\tau_{DM} m_{DM}}
\int_{l.o.s.} \rho_{halo} (\vec{l}) d\vec{l} \; .
\end{equation}
The ratio $ A_{halo}/A_{eg} $ only depends on  cosmological
parameters and the halo dark matter density integrated along the 
line of sight
\footnote{The coefficients $ A_{halo} $ and $A_{eg} $ 
are related to the coefficients $C_\gamma$ and $D_\gamma$
in our previous paper \cite{bcx07} by a factor
$2/m_{3/2}$. Let us also note that in that reference there is
a typo in the definition of $C_\gamma$: it should say $10^{-7}$ 
instead of $10^{-6}$.}. Hence, the intensity and angular distribution of the 
halo signal is very sensitive to the distribution of the dark 
matter in the Milky Way. Surprisingly, for typical halo models, 
this ratio is of order unity~\cite{ahx98}.

Consider a Navarro-Frenk-White profile for the DM matter of our
galaxy, 
\begin{equation}
\rho_{NFW} (r) = \frac{\rho_h}{r/r_c (1+r/r_c)^2}
\label{NFWprofile}
\end{equation}
with $\rho_h = 0.33\ \mbox{GeV/cm$^3$} = 0.6\times 10^5 \rho_c $ 
giving the halo density normalisation and $r_c = 20 $ kpc the critical 
radius where the profile slope changes. For any given point along the 
line of sight, the distance $r$ from the centre of the galaxy can be 
expressed as function of the galactic coordinates, the longitude $l$ 
and latitude $b$, and the distance from the Sun $s$ in units of 
$R_{\odot}$ as
\begin{equation}\label{galcoord}
r^2(s,b,l) = R_{\odot}^2
\left[ (s- \cos b\cos l)^2 + (1- \cos^2 b\cos^2 l)\right]\;.
\end{equation}
The flux factor then reads
\begin{eqnarray}
A_{halo} (b,l) = \frac{R_{\odot}}{4\pi\tau_{DM} m_{DM}}
%\frac{\rho_h r_c}{\rho_c} 
\int_0^{\infty} ds\ \rho_{NFW} (r(s,b,l))\; .
%\frac{ds}
%{\sqrt{ (s- \cos b\cos l)^2 + (1- \cos^2 b\cos^2 l)}}
%\nonumber\\
%&& \times \frac{1}{
%\left(1 + R_{Sun}/r_c \sqrt{ (s- \cos b\cos l)^2 + (1- \cos^2 b\cos^2 l)}
%\right)^2}
% \; .
\end{eqnarray}
\begin{figure}[t]
\begin{center}
\epsfig{file=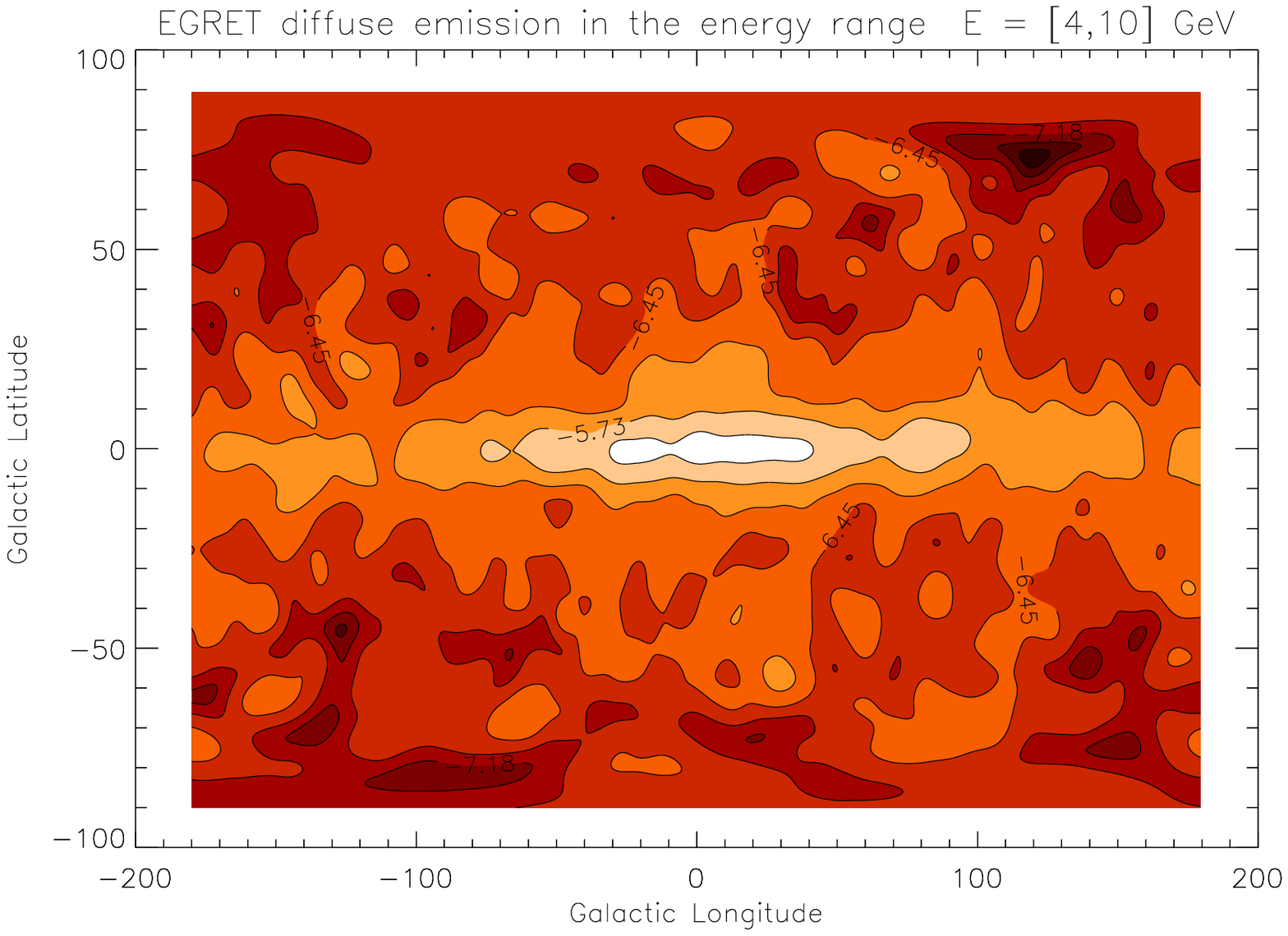,width=0.49\textwidth}
\epsfig{file=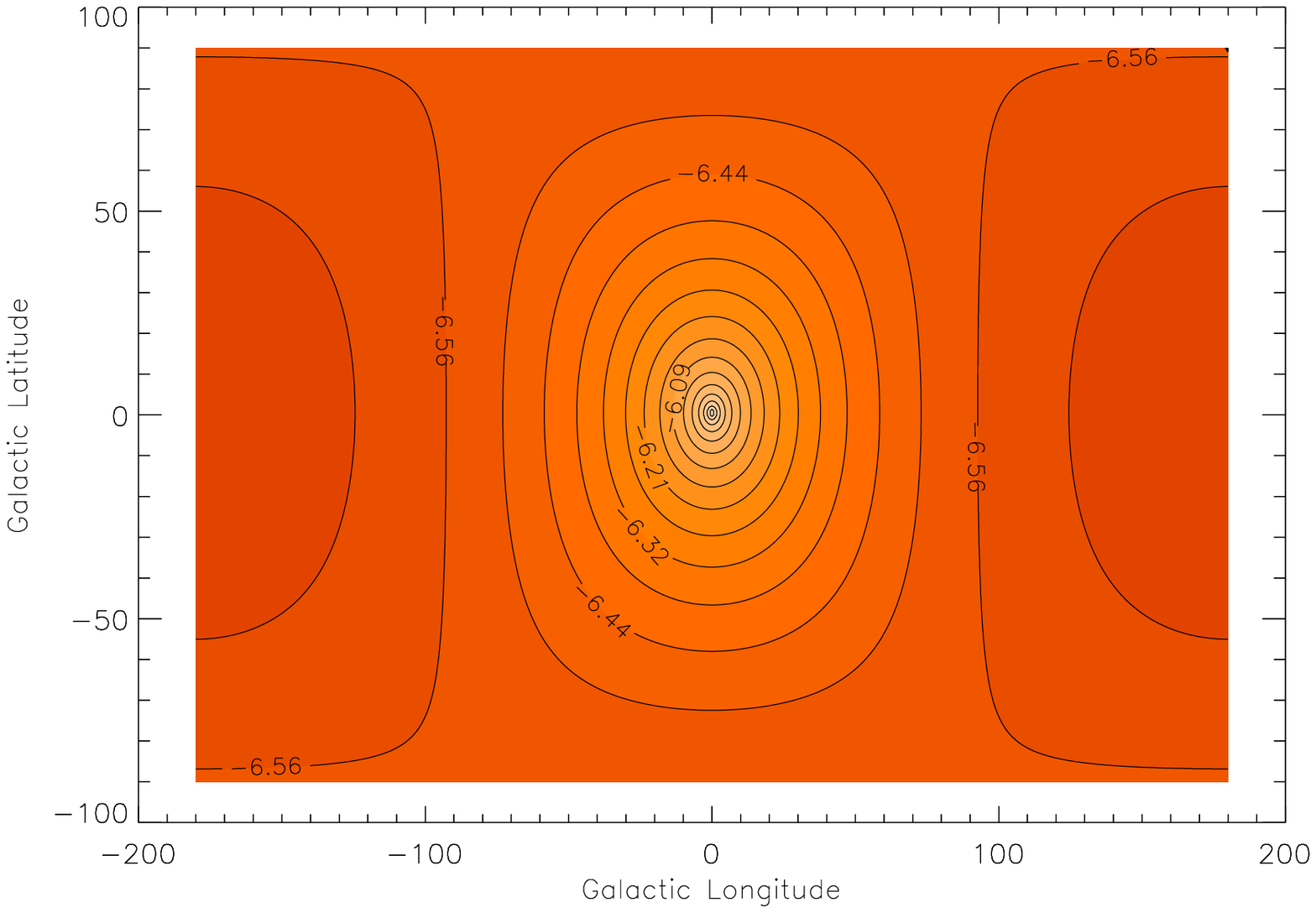,width=0.49\textwidth}
\end{center}  
\caption{
{\it Left:} EGRET diffuse emission in the energy range E=[4,10] GeV. 
{\it Right:} Sum of the Galactic plus extra-galactic contributions 
to the gamma-ray flux from gravitino decay, 
for $\tau_{DM}=4 \times 10^{27}$~s and $m_{DM}=10$~GeV
and the NFW profile given in Eq.~(\ref{NFWprofile}).}
\label{fig:egret}
\end{figure}
This expression can be used to give the flux dependence on the
angle, as long as we are far away from the central cusp. 
In that region a more appropriate quantity is the average flux on the
solid angle corresponding to the detector resolution around the
direction $(b,l)$, i.e.
\begin{equation}
\langle A_{halo} (b,l) \rangle_{\Delta\Omega} = 
\frac{1}{\Delta\Omega} 
\int_{\Delta\Omega} d\Omega\ A_{halo} (b,l)\; ,
\label{eq:finiteangle}
\end{equation}
where the infinitesimal solid angle is given by
$ d\Omega = dl \cos(b) db $.
We consider therefore the average flux for an angular resolution of
$1^\circ$ as measured by EGRET. The result of the numerical 
integration for the halo flux plus the isotropic extragalactic 
component is shown in the right panel of Fig.~\ref{fig:egret}, 
which illustrates
the decrease of photon flux away from the galactic centre, both in 
longitude and latitude. Note that the dependence is not very strong 
and indeed $A_{halo} $ changes only by a factor 20 between the 
galactic centre and the anti-centre, and within a factor of 8
if one cuts out the region within $10$ degrees around the 
galactic plane. Averaging over all sky excluding the galactic
plane, we obtain $\overline{A}_{halo}/A_{eg} \simeq 0.76 $,
so that the line is actually dominating the signal. 
In fact the total flux of diffuse gamma-rays for all sky directions
is given by
\begin{eqnarray}
\Phi_{diff} &=& 
\int d\Omega \int_0^{\infty} dE \left( \frac{d J_{eg}}{dE} 
+ \frac{d J_{halo}}{dE} (b,l) \right)\\
&=& 4\pi A_{eg} \frac{2 \sinh^{-1}(\sqrt{\kappa})}{3 \sqrt{\kappa}} 
+ \int_{-\pi/2}^{\pi/2} \cos(b) db \int_{-\pi}^{\pi}dl A_{halo} (b,l) \\
&\simeq& 4 \pi A_{eg} \left( 0.5 + \frac{\overline{A}_{halo}}{A_{eg}} 
\right) \\
&\simeq &  1.5\times 10^{-6}\ ({\rm cm}^2\ {\rm s})^{-1} 
\left(\frac{\tau_{DM}}{ 10^{27} \  \mbox{s}} \right)^{-1}
\left(\frac{m_{DM}}{ 10 \ \mbox{GeV}} \right)^{-1}\; ,
\end{eqnarray}
where we have used $\kappa = 3 $ and Eqs.~(\ref{cgamma}) and
(\ref{dgamma}).

The expected signal from Dark Matter decay in the halo can be compared 
with the diffuse gamma-ray flux observed by EGRET. 
Contour lines of constant flux, with photon energies between 
4 GeV and 10 GeV are shown in the right panel
of Fig.~\ref{fig:egret}. As expected, the smallest flux is 
observed in the directions of the north and south poles, i.e. 
orthogonal to the galactic disk. According to Fig.~\ref{fig:egret}, the 
signal in these directions is larger than in the direction opposite to
the galactic centre. We therefore conclude that the signal from the 
MW halo could be most effectively observed looking away from the 
galactic disk, which generates most of the background, 
in direction of the poles,
in contrast with the strategy usually adopted for the detection of 
the self-annihilating DM signal.

% Spectrum
\begin{figure}[t]
\begin{center}
\epsfig{file=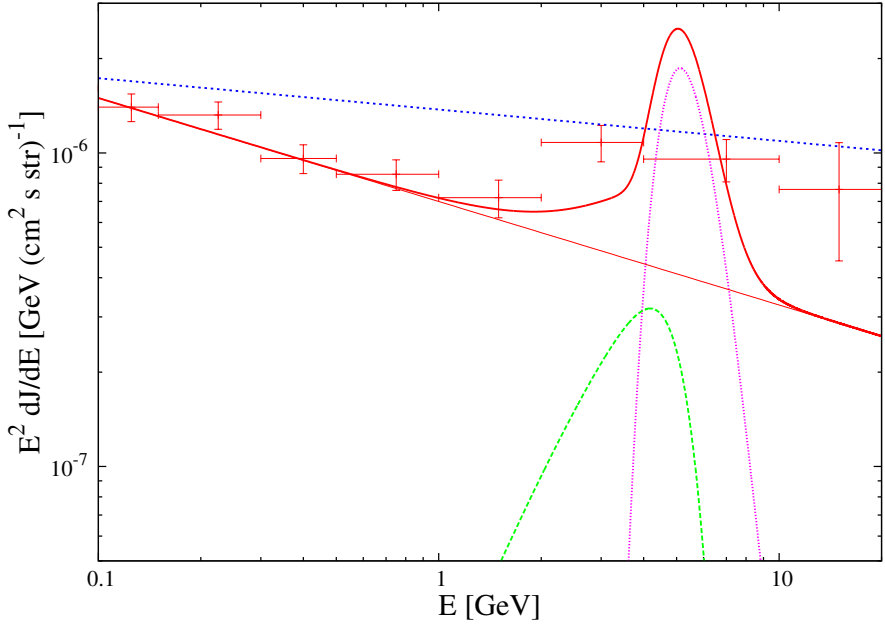,width=12cm}
%,width=12cm,height=6cm}
\end{center}  
\caption{Energy spectrum of extragalactic and halo signal compared to
the EGRET data. The data points are the EGRET extragalactic background
as extracted by Strong {\it et al.} in \cite{Strong:2004ry}, while
the short-dashed (blue) line shows the powerlaw fit 
from Eq.~(\ref{sreekpowerlaw}) obtained previously by 
\cite{Sreekumar:1997un}.
The extragalactic and halo signals for $ \tau_{DM} = 10^{27}\ {\rm s} $ 
and $ m_{DM} = 10\ {\rm GeV} $ are respectively the 
long-dashed (green) and dotted (magenta) lines, while the solid 
(thick red) line shows the sum of these contributions with 
a powerlaw background (thin red line), which has been obtained 
fitting the low energy EGRET points.
}
\label{fig-spectrum}
\end{figure}
The photon spectrum is dominated by the sharp line coming from our 
local halo, while the red-shifted extragalactic signal is 
appreciably lower. The position of the line allows a direct measurement 
of the DM particle mass and the height is inversely proportional to the 
lifetime. In Fig.~\ref{fig-spectrum}, we show the expected signal for 
decaying particle mass of $10 $ GeV and for a lifetime of $10^{27} $ s 
in comparison to the extragalactic EGRET data 
\cite{Sreekumar:1997un, Strong:2004ry}.
We mimic the finite energy resolution of the detector by convolving
the signal with a Gaussian distribution and average the halo signal
over a cone of $80^\circ$ around the poles. The height and width of the 
line depend as usual on the energy resolution of the detector; 
here we have taken $15\%$ as energy resolution, as quoted by EGRET 
in this energy range.
Note also that the DM signal peaked at $5$~GeV corresponds to the 
expectation of the model of gravitino LSP with R-parity and 
\mbox{B-L} breaking discussed in~\cite{bcx07}.
A word of caution though is in order in the comparison between data
and signal: the EGRET extragalactic background displayed here has
been extracted assuming isotropy, while our halo emission is
mildly anisotropic away from the galactic plane. GLAST is expected to 
provide much better data at these energies, allowing a much more detailed
analysis of the angular and spectral properties of the
diffuse gamma-ray flux.

\section{Alternative targets}

The Milky Way has dwarf galaxies as satellites, which have a large 
mass to light ratio, like Draco and Ursa Minor. One may therefore hope 
that the flux from decaying DM is significantly enhanced in these 
directions in the sky like in the case of 
annihilating DM~\cite{Evans:2003sc}.
In the following we shall study the dependence of the enhancement on the
angular resolution of the detector and the mass of the dark matter 
constituents. For simplicity, we use the isothermal profile,
\begin{equation}\label{isothermal}
\rho_{halo}(r) = \frac{\rho_0}{1 + r^2/r_c^2} \;,
\end{equation}
for which one can easily derive simple analytic expressions for the
photon flux.

Integrating along the direction of sight and using Eqs.~(\ref{haloflux}),
 (\ref{dgamma}) and (\ref{galcoord}), one finds for the
photon flux from decaying dark matter in the Milky Way halo \cite{ahx98},
\begin{eqnarray}
J_{halo} (b,l) &=& \frac{1}{4\pi}\frac{1}{\tau_{DM} m_{DM}}
\frac{\rho_0 r_c^2}{R_{\odot}
\sqrt{1-\cos^2 l \cos^2 b + r_c^2/R_{\odot}^2}}
\nonumber\\
&&
\quad \times\left(\frac{\pi}{2} + \tan^{-1}{\left(
\frac{\cos b \cos l}{\sqrt{1-\cos^2 b \cos^2 l + r_c^2/R_{\odot}^2}} 
\right)}\right)\;.
\end{eqnarray}
With $R_{\odot} \simeq 8.5\ {\rm kpc}$, $r_c \simeq 3.5\ {\rm kpc}$
and $\rho_0 \simeq 1.37~{\rm GeV}~{\rm cm}^{-3}$, 
this yields in the direction of Draco 
($b_D=34^\circ, l_D=86^\circ;\ \cos l_D \cos b_D \simeq 0.06$):
\begin{equation}
J_{halo} (b_D,l_D) 
\simeq 0.8\times 10^{-7}\;  ({\rm cm}^2 {\rm s}~{\rm str})^{-1} 
\left(\frac{\tau_{DM}}{10^{27}\ {\rm s}}\right)^{-1}
\left(\frac{m_{DM}}{10\ {\rm GeV}}\right)^{-1} \;.
\end{equation}
 
The photon flux observed from a distant dwarf galaxy crucially depends on
the angular resolution of the detector. Averaging over a cone with small 
opening angle $\delta$, directed toward the centre of the dwarf galaxy, 
one has
\begin{equation}
J_{dg}(\delta) \simeq \frac{1}{2\tau_{DM} m_{DM}} 
\frac{1}{\Delta\Omega} \int_0^\delta \alpha d\alpha
\int_{r_{min}(\alpha)}^{r_{max}(\alpha)} dr r^2 
\frac{1}{r^2 + \alpha^2 r^2}\ 
\rho_{dg}(r(\alpha))\;;
\end{equation}
here $\Delta\Omega = \pi\delta^2$ is the infinitesimal solid angle,
$r(\alpha) = \sqrt{(d-r)^2 + \alpha^2 r^2}$, $d$ is
the distance to the dwarf galaxy, $\rho_{dg}$ is the isothermal
profile in Eq.~(\ref{isothermal}); we have also taken a finite tidal 
radius $r_m$ into account, which leads to the finite integration domain
given by $r(\alpha)^2 \leq r_m^2$.
Performing the integrations, one obtains in the relevant case where 
$\delta \ll 1$ and $\delta d,r_c \ll r_m$,
\begin{equation}\label{finite}
J_{dg}(\delta) \simeq \frac{\rho_0 r_c}{4\tau_{DM} m_{DM}} 
\left(\frac{2r_c}{r_c + \sqrt{r_c^2 + \delta^2 d^2}} 
- \frac{2 r_c}{\pi r_m} + \ldots \right)\;,
\end{equation}
where terms ${\cal O}(\delta^2)$ and ${\cal O}(\delta d r_c/r_m^2)$
have been neglected. Up to corrections ${\cal O}(r_c/d)$, the numerator 
of the prefactor is precisely the line of sight integral of the 
dark matter profile for $r_m \rightarrow \infty $,
\begin{equation}
\pi \rho_0 r_c = \int_{l.o.s} dr \rho_{dg} (r)\;,
\end{equation} 
and for $\delta, 1/r_m \rightarrow 0$, the bracket becomes one. 

Taking as an example Draco, typical parameters are 
% $v_h = 22\ {\rm km/s}$,
$r_c = 0.1\ {\rm kpc}$, $\rho^{Dr}_0 = 28.4\ {\rm GeV\;cm^{-3}}$,
$r^{Dr}_m = 1.7\ {\rm kpc}$, $d^{Dr} = 80\ {\rm kpc}$ ~\cite{bnx06}. 
The correction due to the finite tidal radius in
Eq.~(\ref{finite}) is then negligible, and in the line of sight 
approximation
($\delta = 0$) one obtains the flux
\begin{equation}\label{JDraco}
J_{D}(0) \simeq 2\times 10^{-7}\;  
({\rm cm}^2 {\rm s}~{\rm str})^{-1} 
\left(\frac{\tau_{DM}}{10^{27}\ {\rm s}}\right)^{-1}
\left(\frac{m_{DM}}{10\ {\rm GeV}}\right)^{-1}\;,
\end{equation}
which is about three times larger than the flux from the halo in the
same direction, as pointed out in \cite{bnx06}. 
The correction factor for finite opening angle in Eq.~(\ref{finite}) 
can only be neglected for $\delta < r_c/d = 1.3\times 10^{-3}$,
which corresponds to the angular resolution of GLAST of about $0.1^\circ$.
For $\delta \simeq 0.004$ the signals from Draco and the halo have equal
strength, but the corresponding field of view is 
$\Delta\Omega \sim 5\times 10^{-5}$, yielding the flux 
$\Phi_{D}(0.004) \sim 3.6\times 10^{-12} ({\rm cm}^2 {\rm s})^{-1} $,
too small to be observed even by GLAST~\cite{GLAST}. 
We conclude that for dark matter particles with masses in the GeV range 
the flux enhancement in the direction of dwarf galaxies is currently 
not of interest. This is different for masses in the keV 
range~\cite{bnx06} where the flux is six orders of magnitude larger 
for the same DM density and in which case the angular resolutions of 
the X-rays detector is much better than for gamma rays.

Another potentially important source of gamma rays is
the Andromeda Galaxy, due to its proximity, large apparent
size and privileged position in the sky away from the Milky 
Way centre. To estimate the photon flux received from the Andromeda
Galaxy, we approximate the dark matter distribution
by the isothermal profile, Eq.~(\ref{isothermal}),
with $r^{\rm M31}_c = 1.5\ {\rm kpc}$, 
$\rho^{\rm M31}_0\simeq 15.7 \ {\rm GeV\;cm^{-3}}$,
$r^{\rm M31}_m = 117\ {\rm kpc}$ ~\cite{Tempel:2007eu}.
The total flux received from a cone directed toward
the centre of the Andromeda Galaxy with a small opening
angle $\delta$ is given by Eq.~({\ref{finite}).
On the other hand, for the angular resolution of GLAST, the gamma ray
flux from the Milky Way halo in the direction to the 
Andromeda Galaxy ($b_{\rm M31}=-22^\circ, l_{\rm M31}=121^\circ$)
is approximately 
\begin{eqnarray}
J_{halo}(b_{\rm M31},l_{\rm M31})\simeq 0.5\times 10^{-7}\;  
({\rm cm}^2 {\rm s}~{\rm str})^{-1} 
\left(\frac{\tau_{DM}}{10^{27}\ {\rm s}}\right)^{-1}
\left(\frac{m_{DM}}{10\ {\rm GeV}}\right)^{-1}\;.
\end{eqnarray}
The opening angle at which the Milky Way signal equals the
Andromeda Galaxy signal is around $5^\circ$, 
which corresponds to a flux 
$\Phi_{\rm M31}(5^\circ)\simeq 1.5\times 10^{-9} ({\rm cm}^2 
{\rm s})^{-1}$. This is above the sensitivity of detection
of GLAST (although below the EGRET sensitivity), hence
we conclude that GLAST {\it might} be able to see
the Andromeda Galaxy as a gamma ray source over the
background from the decaying dark matter in the Milky 
Way halo. However, one should also note that the extraction
of an extragalactic gamma ray flux at such low galactic
latitudes is intricate, and the Andromeda signal from
dark matter decay could be easily masked by photons from
standard astrophysical processes occurring in the Milky Way disk.

Finally, let us discuss other potentially interesting
sources of gamma rays, namely nearby
galaxy clusters located at high galactic latitudes,
like Coma or Virgo. The photon flux can be
computed along the same lines as for dwarf galaxies, where
the density profile Eq.~(\ref{isothermal}) has to be replaced
by the profile for the isothermal $\beta$-model 
\cite{Cavaliere:1976tx,Sarazin:1977}
\begin{equation}\label{beta-model}
\rho^{cl}=\rho^{cl}_0 \frac{3+r^2/r^2_c}{(1+r^2/r^2_c)^2}\;,
\end{equation}
that approximately describes the distribution of dark matter in
a galaxy cluster. The result for $\delta \ll 1$ and 
$\delta d,r_c \ll r_m$ can be approximated by:
\begin{equation}\label{Jclusters}
J_{cl}(\delta)\simeq \frac{\rho^{cl}_0 r_c}{2\tau_{DM} m_{DM}} 
\left(\frac{r_c}{\sqrt{r_c^2 + \delta^2 d^2}} 
- \frac{r_c}{\pi r_m} + \ldots \right)\;.
\end{equation}

Nearby galaxy clusters have a large angular size and
could not appear as point sources. To be precise, the core 
of the Coma Cluster has a size of 0.3 Mpc compared to a 
distance of 98 Mpc, corresponding to $0.17^\circ $ of angular 
size, which is larger than the angular resolution of GLAST.
On the other hand, the Virgo Cluster has a core radius
of 10 kpc and lies at 18 Mpc, which translate into
an angular size of $0.03^\circ$, slightly smaller than
the angular resolution of GLAST. In consequence, for these
objects the line of sight approximation $\delta=0$ is
not a good approximation and in order to obtain a reliable
estimate for the photon flux the complete expression 
Eq.~(\ref{Jclusters}) has to be used. Taking 
$\rho^{cl}_0=10^{-2}\ {\rm GeV\;cm^{-3}}$ and
$\rho^{cl}_0=2.3\ {\rm GeV\;cm^{-3}}$ for the Coma and
Virgo clusters, respectively, we obtain for the angular
resolution of GLAST 
\begin{eqnarray}
J_{\rm Coma}(0.1^\circ)\simeq 4\times 10^{-7}\;  
({\rm cm}^2 {\rm s}~{\rm str})^{-1} 
\left(\frac{\tau_{DM}}{10^{27}\ {\rm s}}\right)^{-1}
\left(\frac{m_{DM}}{10\ {\rm GeV}}\right)^{-1}\;,\nonumber \\
J_{\rm Virgo}(0.1^\circ)\simeq 10^{-6}\;  
({\rm cm}^2 {\rm s}~{\rm str})^{-1} 
\left(\frac{\tau_{DM}}{10^{27}\ {\rm s}}\right)^{-1}
\left(\frac{m_{DM}}{10\ {\rm GeV}}\right)^{-1}\;.
\end{eqnarray}
which are around one order of magnitude larger than 
the photon flux from the Milky Way halo in the direction
of the galactic pole, 
\begin{equation}
J_{halo} (b=\pi/2) 
\simeq 0.7\times 10^{-7}\;  ({\rm cm}^2 {\rm s}~{\rm str})^{-1} 
\left(\frac{\tau_{DM}}{10^{27}\ {\rm s}}\right)^{-1}
\left(\frac{m_{DM}}{10\ {\rm GeV}}\right)^{-1} \;.
\end{equation}
Nevertheless, the total flux received from these objects
is so small that in order to distinguish a gamma ray
signal from the Coma and Virgo clusters, it would be
necessary a sensitivity  of $7\times 10^{-11}$ 
photons ${\rm cm}^{-2}\;{\rm s}^{-1}$, which is more than
one order of magnitude lower than the GLAST sensitivity~\cite{GLAST}. 

\section{Discussion}

Should the gamma-ray line from DM decay be discovered at high galactic 
latitude, the problem will arise of how to discriminate it from 
scenarios with self-annihilating DM. The presence of a spectral line 
in the diffuse signal at high galactic latitude has in fact been 
proposed, several years ago, as a signature of annihilating DM particles
~\cite{Bergstrom:2001jj} (see also 
Refs.~\cite{Taylor:2002zd,Ullio:2002pj}). 
We note however that there are at least four important differences 
between the two scenarios:
\begin{itemize}
\item {\it the angular profile} of the predicted gamma-ray signal; 
\item {\it the comparison between galactic and extra-galactic 
components;}
\item {\it the ratio between the line and the gamma-ray continuum;}
\item {\it the angular power spectrum} of the gamma-ray background.
\end{itemize}

\begin{figure}[t]
\begin{center}
\epsfig{file=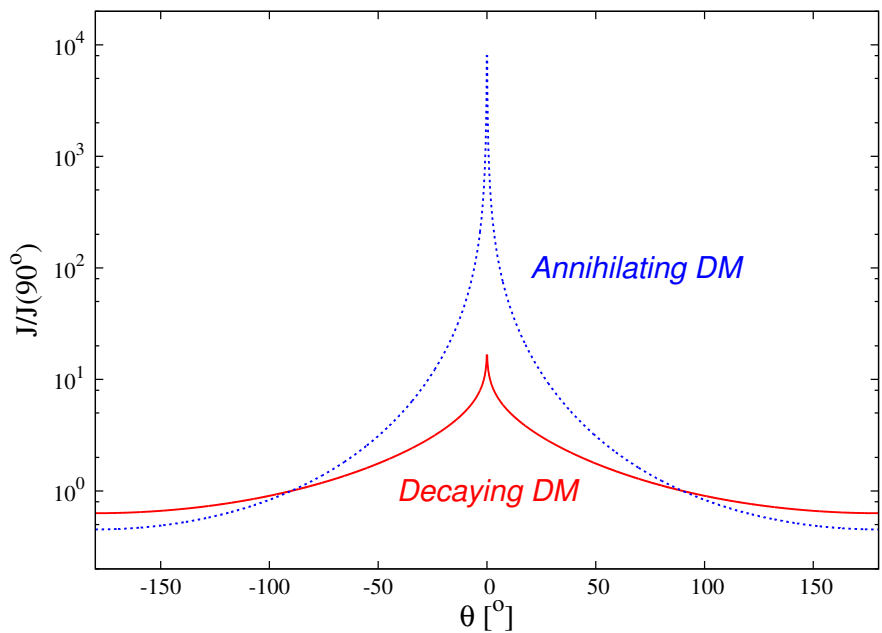,width=12cm}
%,width=12cm,height=6cm}
\end{center}  
\caption{Angular profile of the gamma-ray signal as function of 
the angle $\theta $ to the centre of the galaxy for a NFW 
halo distribution for decaying DM, solid (red) line, 
compared to the case of self-annihilating DM, dashed (blue) line.
Both signals have been normalised to their values
at the galactic poles, $\theta = \pm 90^\circ$.
%The self-annihilating signal has been normalised to be equal to
%the decaying one at the galactic poles, $\theta = \pm 90^\circ$.
The central cusp is regularised by assuming in both cases the
GLAST angular resolution of $0.1^\circ$ and integrating on the solid 
angle as in Eq.~(\ref{eq:finiteangle}).}
\label{fig:angular}
\end{figure}

We stress in fact that in the case of self-annihilating DM, 
the gamma-ray flux from a given direction in the sky is a steeply 
falling function of the angle with respect to the Galactic centre, 
and it is in particular much steeper than the flux from decaying DM, 
as can be seen in Fig.~\ref{fig:angular}. That is precisely why 
we have focused on high galactic latitudes in Sec. 2. 
Furthermore, the extra-galactic component of the gamma-ray flux in the 
case of annihilating DM is unlikely to be detected in absence of a 
strong signal from the Galactic centre~\cite{Ando:2005hr}.
In other words, in order for an extra-galactic component to be detected 
at high latitudes, the gamma-ray signal from the galactic centre should 
be easily detectable, despite the much stronger astrophysical background 
toward the innermost regions of the Galaxy. Thus, in case a line is 
observed at high galactic latitude, the absence of a similar line 
in the gamma-ray spectrum of the Galactic centre would favour an 
interpretation in terms of decaying DM.

It is still possible that astrophysical processes, such as the 
formation of spikes~\cite{Gondolo:1999ef,Bertone:2002je,Bertone:2005hw} 
and mini-spikes~\cite{Bertone:2005xz,Bertone:2006nq,Zhao:2005zr}, 
i.e. large DM overdensities around Supermassive and Intermediate Mass
Black Holes respectively, modify this picture, by boosting the
gamma-ray signal from cosmological structures much more than 
the flux from the Galactic centre~\cite{Ahn:2007ty,Horiuchi:2006de}.
Even in this case, however, it should be possible to discriminate
between the two scenarios, by studying the spectral features of 
the extra-galactic background. In fact, for most self-annihilating
DM candidates, such as the supersymmetric neutralino and the
B$^{(1)}$ particle in theories with Universal Extra Dimensions,
direct annihilation to photons is severely suppressed with respect 
to other channels such as annihilation to quarks or gauge bosons. 
It follows that the annihilation spectra are characterised by a 
continuum emission that is inevitably associated with the line signal, 
as shown in Fig.~\ref{fig:spectra}.
Furthermore, masses below 50 GeV are usually considered unlikely
for annihilating candidates. For instance, the current constraint
in the neutralino mass is 40 GeV (assuming unification of gaugino
mass parameters at the GUT scale)~\cite{Yao:2006px}
while electroweak precision data exclude B$^{(1)}$ masses below 
300 GeV~\cite{Gogoladze:2006br}. 

\begin{figure}[t]
\begin{center}
\epsfig{file=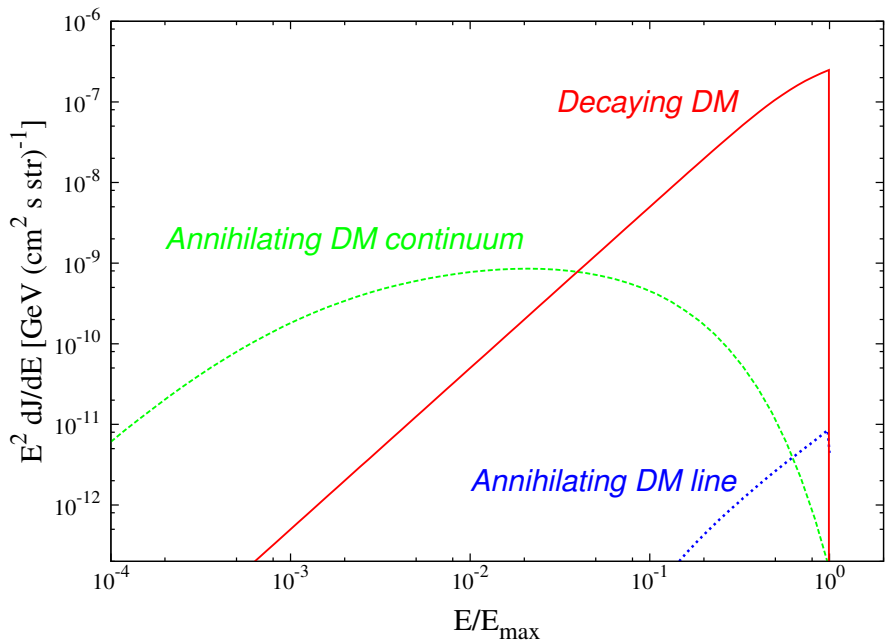,width=12cm}
%,width=12cm,height=6cm
\end{center}  
\caption{Spectrum of decaying DM extra-galactic component, 
solid (red) lines, 
compared with the spectrum of annihilating DM 
(continuum as long-dashed (green) plus line 
in short-dashed (blue)). For the case of decaying
DM, $E_{max}=m_{DM}/2$, whereas for the case
of annihilating DM, $E_{max}=m_{DM}$.}
\label{fig:spectra}
\end{figure}

Although the relative importance of the line can be in some cases 
particularly high, as e.g. in the case of Inert Higgs 
DM ~\cite{Gustafsson:2007pc}, the peculiar shape of the angular 
power spectrum of the gamma-ray background~\cite{Ando:2005xg}
can be used as a diagnostic tool to discriminate between 
annihilating and decaying DM. As for all the other strategies,
this method heavily relies on the assumption that a sufficient
number of photons are collected above the astrophysical 
background to allow a statistically meaningful analysis. 
In the worst case scenario, null GLAST searches can be used to 
exclude regions of the parameter space that lead to observable fluxes.

In Fig.~\ref{fig:exclusion} we show an exclusion plot in the 
($m_{DM},\tau_{DM}$) plane, where we show the regions of the parameter 
space that already are ruled out by a comparison with EGRET data. 
In order to investigate the discovery potential of GLAST, we refer to 
the official publications of the collaboration \cite{GLAST}
where the sensitivity 
to astrophysical lines as a function of energy is studied for two 
different sky regions, namely an 'annulus' region where the signal 
to background ratio is maximised for annihilating DM candidates, 
and a 'high latitude' region, defined as the region of the sky with 
galactic latitude higher than 20 degrees and excluding a 35 degrees 
circle around the Galactic centre. In our case, there is a considerable 
reduction of the background, and a not-so-large decrease of the signal 
when going at high galactic latitude, for the reasons discussed above. 
So it is not a surprise that the 'high latitude' region is more
favourable for constraining decaying DM.
In Fig.~\ref{fig:exclusion} we show the reach of GLAST for the
two regions along with the EGRET-excluded region.
For the present bound we take conservatively the requirement that the 
peak in the energy spectrum with $15\%$ energy resolution
remains below the $2\sigma $ band of the EGRET spectrum obtained by 
Sreekumar {\it et al.} in~\cite{Sreekumar:1997un},
\begin{equation}\label{sreekpowerlaw}
\frac{dJ_{\sc EGRET}}{dE} =
(7.32 \pm 0.34) \times 10^{-6}  
({\rm cm}^2~{\rm s}~{\rm str}\  {\rm GeV})^{-1} 
\left(\frac{E}{0.451 \ {\rm GeV}}\right)^{-2.1\pm 0.03}\; .
\end{equation}
The different mass dependence of the EGRET bound is due to the fact 
that it is a constraint on $\frac{dJ}{dE} $ instead than 
on the integrated flux and on the energy dependence of the GLAST
sensitivity.
The limit disappears at $100-120 $ GeV, which is the maximal energy 
plotted in the EGRET fit, but note that data above 10 GeV have large 
errors and were not used to obtain Eq.~(\ref{sreekpowerlaw}).

\begin{figure}
\begin{center}
\epsfig{file=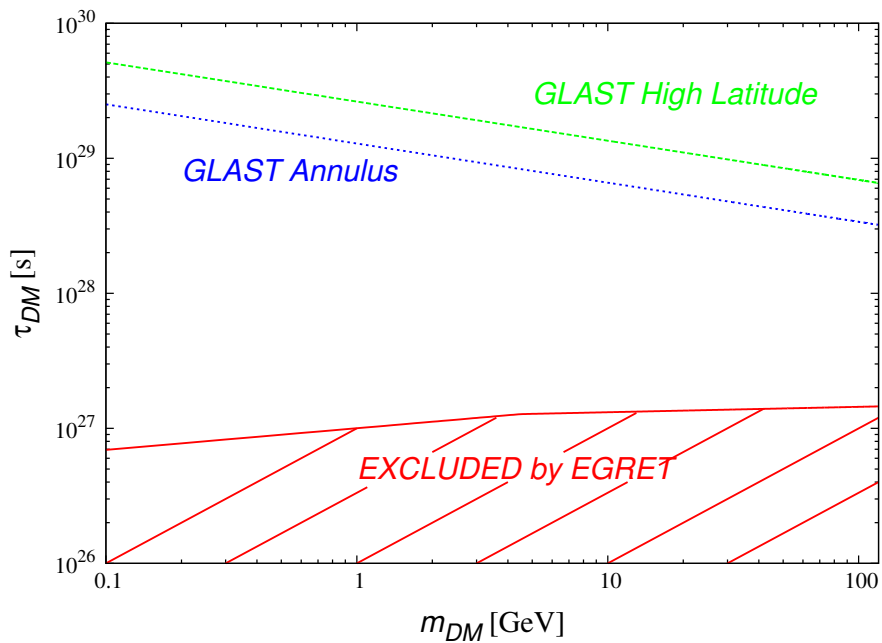,width=12cm}
%,width=10cm,height=6cm}
\end{center}  
\caption{Exclusion plot in the ($m_{DM},\tau_{DM}$) plane based on 
EGRET data. All the region below the solid (red) line is excluded
by the requirement that the halo line is below the measured
flux as explained in the text. We also show the region of the 
parameter space where GLAST could discover the annihilation line:
the short-dashed (blue) line corresponds to the GLAST sensitivity 
in the 'annulus' region, while the long-dashed (green) line to the 
'high latitude' region.
}
\label{fig:exclusion}
\end{figure}

\section{Conclusions}

We have discussed the prospects for indirect detection of gravitino
DM in R-parity breaking vacua with upcoming experiments such as
GLAST and AMS-02. The search strategy is in this case significantly 
different with respect to annihilating DM particles, due to different 
angular profile of the predicted gamma-ray signal and the different 
relative importance of the extra-galactic background. We found that the 
predicted signal from DM decays {\it in the galactic halo} would resemble 
an extra-galactic component with a shape and normalisation similar 
to the one inferred by EGRET data.

Without trying to fit the controversial EGRET data with our model,
we have determined the regions of the ($m_{DM},\tau_{DM}$) plane where 
GLAST could discover the gamma-ray line from gravitino decay,
and we discussed how to discriminate this signal from astrophysical
sources and from a signal originating from annihilating DM.
This discrimination is based on the fact that: the line
may appear at energies below 50 GeV, which is somewhat challenging for 
popular annihilating DM candidates; it would exhibit no continuum 
flux (unless the mass of the gravitino is above the W mass);
there would be a weak line signal from the galactic centre, possibly
hidden in the galactic background; the angular power spectrum of the 
signal would be different from the case of annihilating DM, and 
from extra-galactic astrophysical sources.

In the worst case scenario, i.e. in case of null GLAST searches, 
the data can be used to set constraints in the ($m_{DM},\tau_{DM}$)
plane and improve the bound on the DM lifetime into photons
by more than one order of magnitude, actually even
more than two for masses below 1~GeV.

\section*{Acknowledgements}

We would like to thank to E.-J.~Ahn, K.~Hamaguchi, G.~Heinzelmann, 
J.~Ripken, D.~Tran and T.~T.~Yanagida for valuable discussions, 
and J.~Conrad for useful comments on the manuscript.
We also thank the CERN Theory Institute on the Interplay between 
Cosmology and LHC for hospitality during the completion of this paper.
GB and LC acknowledge the support of the ``Impuls- and Vernetzungsfond''
of the Helmholtz Association, contract number VH-NG-006.

\end{document}